\documentclass[aps,prl,twocolumn,superscriptaddress,showpacs]{revtex4}
\usepackage{graphicx}
\usepackage{amsmath}

\begin{document}

\title{Effect of adhesion geometry and rigidity on cellular force distributions}
\author{Ilka B. Bischofs}
\email[]{ilka@bischofs-pfeifer.eu}
\affiliation{University of Heidelberg, Bioquant, Im Neuenheimer Feld 267, 69120 Heidelberg, Germany}
\affiliation{Lawrence Berkeley National Lab, Physical Biosciences Division, Berkeley, CA 94710, USA}
\author{Sebastian S. Schmidt}
\affiliation{University of Heidelberg, Bioquant, Im Neuenheimer Feld 267, 69120 Heidelberg, Germany}
\affiliation{Helmholtz-Zentrum Berlin, Glienicker Strasse 100, 14109 Berlin, Germany}
\author{Ulrich S. Schwarz}
\email[]{Ulrich.Schwarz@kit.edu}
\affiliation{University of Heidelberg, Bioquant, Im Neuenheimer Feld 267, 69120 Heidelberg, Germany}
\affiliation{University of Karlsruhe, Theoretical Biophysics Group, 76128 Karlsruhe, Germany}

\date{\today}

\begin{abstract}
  The behaviour and fate of tissue cells is controlled by the rigidity
  and geometry of their adhesive environment, possibly through forces
  localized to sites of adhesion. We introduce a mechanical model that
  predicts cellular force distributions for cells adhering to adhesive
  patterns with different geometries and rigidities. For continuous
  adhesion along a closed contour, forces are predicted to be
  localized to the corners. For discrete sites of adhesion, the model
  predicts the forces to be mainly determined by the lateral pull of
  the cell contour. With increasing distance between two neighboring
  sites of adhesion, the adhesion force increases because cell shape
  results in steeper pulling directions. Softer substrates result in
  smaller forces. Our predictions agree well with experimental force
  patterns measured on pillar assays.
\end{abstract}



\maketitle

Adherent tissue cells react very sensitively to the physical
properties of their environment, including mechanical stiffness and
the spatial distribution of adhesive cues \cite{c:geig09}. On flat
substrates, mechanical stiffness and adhesive geometry can be
controlled by using soft elastic substrates \cite{c:pelh97} and
microcontact printing of adhesive islands \cite{c:chen97},
respectively. Rigidity and geometry can be altered simultaneously
combining the above techniques \cite{c:wang02,c:goff06} or by using
bio-functionalized pillar assays, in which cells adhere to the tops of
an array of flexible micro-needles \cite{c:lemm05,c:saez07}.  Such
biophysical approaches have revealed that stiffness and geometry
sensing are closely related as they both involve forces being
localized at discrete adhesion sites.

Despite the tremendous experimental progress in this field, our
theoretical understanding of the relation between stiffness sensing,
geometry sensing and cellular force distributions is still very
limited.  Here we analyze a simple theoretical model that describes
the relation between cell shape and force distribution.  The shape of
tissue cells adhering to discrete sites of adhesion is dominated by
the formation of inward curved circular arcs of the non-adherent
contour \cite{c:zand89,c:barz99,c:ther06,uss:bisc08a}.  This shape
feature follows from a modified Laplace law of competing effective
surface tension and effective line tension. Based on these concepts we
calculate traction forces of stationary cells as a function of
adhesive geometry and stiffness.

\begin{figure}
\includegraphics[width=\columnwidth]{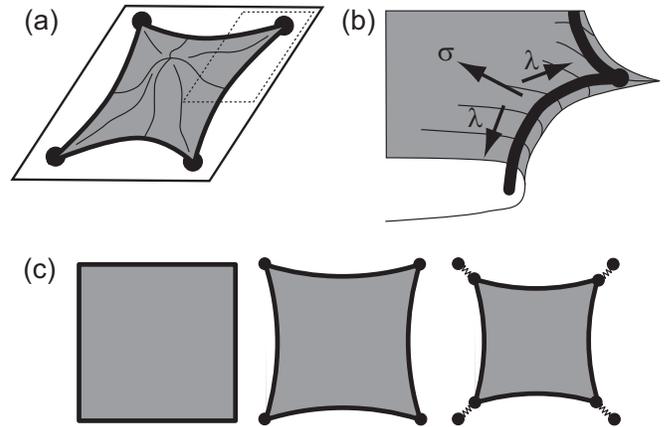}
\caption{\label{Cartoon}(a) Cartoon of a typical tissue cell adhering
  to a flat substrate at four discrete sites of adhesion.  In
  (b) we focus on the framed region. (b) Physical forces acting at the
  free boundary between two neighboring sites of adhesion: while the
  surface tension $\sigma$ pulls the contour inward, the line tension $\lambda$ pulls the
  contour straight. The black line represents an actin cable
  reinforcing the contour. (c) Here we consider three
  cases of increasing complexity: cells adhering to one large adhesive
  island, cells adhering to discrete sites of adhesion on a rigid
  substrate, and cells adhering to discrete sites of adhesion on a
  soft substrate.}
\end{figure}

\textit{Model.} Fully spread tissue cells typically flatten with only
the nucleus sticking out, as shown schematically in
Fig.~\ref{Cartoon}a. Often they adhere to the substrate at
sites of adhesion distributed along the cell boundary.  In this case
it is appropriate to consider an effectively two-dimensional (2D)
model which parameterizes cell shape by its 2D contour $\vec r(l)$
with the internal coordinate $l$. After cell spreading is completed
the cellular forces are mainly contractile.  Fig.~\ref{Cartoon}b
depicts the physical forces acting at the cell boundary
\cite{c:zand89}.  First the cell contour is drawn inward mainly due to
spatially distributed tension in the actin cytoskeleton and the plasma
membrane.  In the 2D model, this effect is described by an effective
surface tension $\sigma$, because the main effect is to reduce 2D
surface area. Second the cell contour resists the inward pull. Often
the cell contour is reinforced by the assembly of contractile actin
cables connecting neighboring sites of adhesion. In the 2D model, this
effect is described by an effective line tension $\lambda$ which
contracts the contour between adhesion sites. In mechanical
equilibrium the contractile forces exerted by cells are primarily
balanced on the adhesive substrate \cite{c:wang01a}.  To derive the
force distribution for a given adhesive geometry we therefore have to
minimize the effective energy functional for the cell contour under
the appropriate adhesion constraint:
\begin{equation}
E = \int \sigma dA + \int \lambda dl + \int \vec f_{ad} \cdot (\vec r -\vec r_0) dl.
\label{energy}
\end{equation}
The first integral extends over the cell surface area $A$ and the
second and third along the cell contour $l$. The third term introduces
the adhesion tension $\vec f_{ad}$ as a Lagrangian parameter for the
constraint set by the adhesion geometry and described by $\vec r_0$.
Based on this model we calculate the cellular adhesion forces by
solving the corresponding Euler-Lagrange equation for the three cases
depicted in Fig.~\ref{Cartoon}c.

\begin{figure}
\includegraphics[width=\columnwidth]{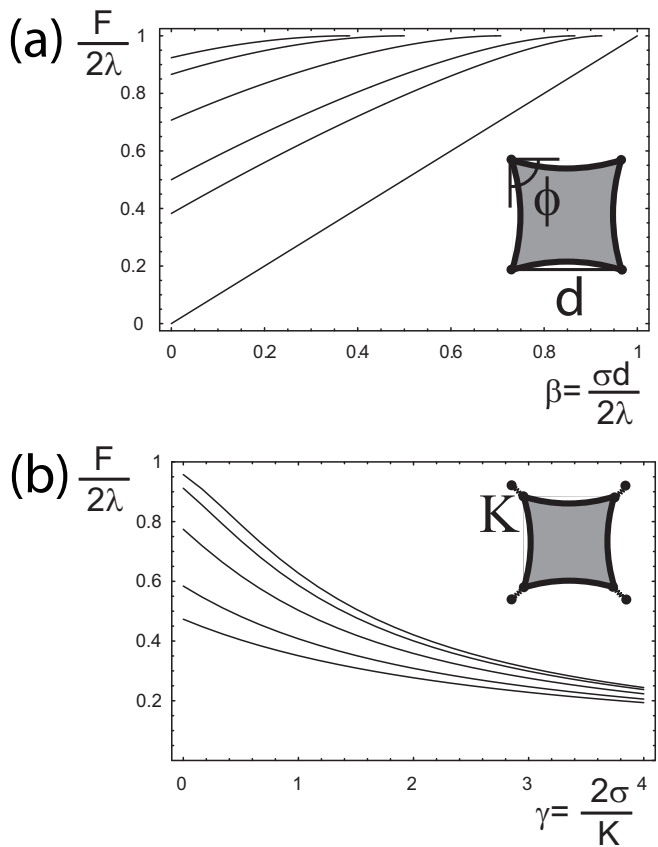}
\caption{\label{Plots}(a) Dimensionless force $F/2 \lambda$ at equally
  spaced adhesion points as a function of the dimensionless spanning distance $\beta =
  \sigma d / 2 \lambda$ for different opening angles $\phi = \pi/4,
  \pi/3, \pi/2, 2\pi/3, 3\pi/4, \pi$ from top to bottom.  (b)
  Dimensionless force $F/2 \lambda$ as a function of the
  dimensionless inverse stiffness $\gamma = 2 \sigma / K$ for adhesion to regular polygons
  with $\phi = \pi/4, \pi/3, \pi/2, 2\pi/3, 3\pi/4$ from top to bottom and $\beta = 0.1$.}
\end{figure}

\textit{Continuous adhesion.} For cells that adhere continuously to
the substrate, the cellular shape $\vec r$ is fixed to $\vec r_0$.
The adhesion tension acting along the contour then follows from
Eq.~\ref{energy}:
\begin{equation}
\vec f_{ad} = -[\sigma + \lambda \kappa ] \vec n
\label{adhesionforce}
\end{equation}
where $\vec n$ is the normal vector of the contour curve and $\kappa$
its local curvature. For convex shapes ($\kappa > 0$), surface and
line tension conspire to pull the cell contour inward, while for
concave shapes ($\kappa < 0$), the line tension opposes the surface
tension, thereby decreasing the adhesion force. For example, a cell
adhering to a circular patch of radius $R$ applies a constant and
inwardly directed adhesive tension $\sigma + \lambda / R$ along the
contour. Along straight boundaries ($\kappa = 0$), e.g.\ along the
straight lines of polygonal cells, the contribution of the line
tension $\lambda$ vanishes and the adhesion tension is simply
$\sigma$.  The contribution of the line tension localizes to the
corners of the polygon.  Approximating a corner with opening angle
$\phi$ by an arc with radius $\epsilon$ and then taking the limit to a
sharp corner by $\epsilon \rightarrow 0$, we can calculate the
adhesion force $\vec F$ acting in the corner:
\begin{equation}
\vec F = \lim_{\epsilon\rightarrow0}\int_{-\frac{\varphi}{2}}^{\frac{\varphi}{2}}
(\sigma+\frac{\lambda}{\epsilon})\vec n(\theta) \epsilon d\theta
= 2 \lambda \cos{\left( \frac{\phi}{2} \right)} \vec n_{b}
\label{corner}
\end{equation}
where $\varphi = \pi-\phi$ and $\vec n_{b}$ points in the direction of
the bisecting line. The model predicts that the smaller $\phi$, the
larger the force pulling on the corner, with a maximal value of $2
\lambda$ when both arcs pull in the same direction. When considering a
finite radius $r$ of the adhesion, one has to set $\epsilon = r$
rather than taking the limit $\epsilon \rightarrow 0$, resulting in an
additional contribution by the surface tension $\sigma$.
Experimentally cells have been forced into various shapes and forces
have been measured by using adhesive islands on compliant substrates
\cite{c:wang02}. On polygonal islands
strong traction forces are measured at the corners. Our model
qualitatively predicts this corner effect and makes quantitative
predictions on the scaling with $\phi$ which can be experimentally
tested in the future.

\textit{Discrete adhesion.} For cells that adhere to the substrate at
discrete points of adhesion, the cell contour between adhesion points
is free and the shape equation resulting from Eq.~\ref{energy}
predicts the formation of circular arcs with curvature $\kappa =
\sigma / \lambda$. The adhesion force follows from Eq.~\ref{corner} by
replacing the opening angle $\phi$ with the actual pulling angle
$\phi^{\star}\neq\phi$ spanned by the two contour arcs pulling on the
adhesion site. $\phi^{\star}$ has to be derived from the shape
equation.  For three equally spaced adhesion sites with distance $d$
spanning an angle $\phi$ we can derive an explicit equation for the
resultant force on the central adhesion site as a function of the
adhesion geometry:
\begin{equation}
\vec F = 2 \lambda \left[ \beta \sin \left( \frac{\phi}{2} \right)
+ \sqrt{1-\beta^2} \cos \left( \frac{\phi}{2} \right) \right] \vec n_{b}
\label{arc}
\end{equation}
with $\beta = \sigma d / 2 \lambda$ being a dimensionless measure for
the strength of the inward pull. $\beta$ can also be interpreted as
dimensionless spanning distance $d$. Hence, the force scales again
with the line tension $\lambda$ but now also depends on the spanning
distance $d$ and the surface tension $\sigma$. Fig.~\ref{Plots}a
demonstrates that the larger the spanning distance $d$ and the more
acute the opening angle $\phi$, the steeper the inward pull and the
closer the force comes to its maximal value $2 \lambda$. This maximal
value is reached at a critical pulling strength $\beta_c =
\sin(\phi/2)$ when the pulling direction has become so steep that the
two arcs physically touch each other $(\phi^{\star}=0)$, at which
point other physical processes will take over (e.g.  pearling of
tubular extensions \cite{c:barz99}). For equally spaced adhesion sites
along a straight line $\phi =\pi$ the adhesion force scales exactly
linear with spanning distance $d$.  As the opening angle $\phi$
increases beyond $\pi$, the traction force may become a pushing force.
For unequal spacing $d_1 \neq d_2$ between adhesion sites, a
generalization of Eq.~\ref{arc} can be derived, which shows that the
force direction is now tilted towards the side with larger spanning
distance.

\begin{figure}
\includegraphics[width=\columnwidth]{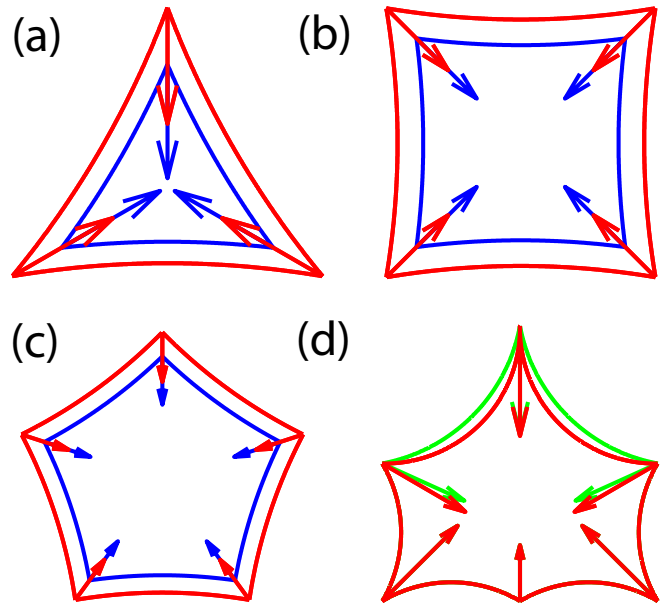}
\caption{\label{Shapes}(a-c) Predicted cell shapes and adhesion forces
  for various adhesion geometries on stiff $\gamma=0$ (red) and
  compliant substrates $\gamma = 1/15$ (blue). Parameters $\beta =
  1/6$ and $\phi=\pi/3,\pi/2$ and $3/5 \pi$ from a to c.  (d)
  Predicted cell shape and adhesion forces on a rigid substrate using
  constant line tension (red) and the tension elasticity model (green,
  $\alpha = 1$, $l_f = 20a$) for an adhesion geometry with different
  spanning distances, namely $d=a$ at the sides and $d=\sqrt 2 a$ at
  the top. In the constant tension model $\lambda$ was adjusted to
  result in the same force for $d=a$.  }
\end{figure}

\textit{Elastic substrate.} To study how in our model elasticity
affects the adhesion force, we now consider discrete adhesion sites
that move in a harmonic potential with spring constant $K$. For
simplicity, we only consider regular polygons, for which we find:
\begin{equation}
\vec F = 2 \lambda \left[ \frac{ \beta \gamma_{\phi} + 
\cos \left( \frac{\phi}{2} \right) \sqrt{\gamma_{\phi}^2 +
\cos^2 \left( \frac{\phi}{2} \right) - \beta^2}}{\gamma_{\phi}^2+
\cos^2 \left( \frac{\phi}{2} \right) } \right] \vec n_{b}
\label{elasticsubstrate}
\end{equation}
with $\gamma_{\phi} = \sin(\phi/2) + \gamma \cos(\phi/2)$. Here
$\gamma = 2 \sigma / K$ is a dimensionless inverse stiffness.
Fig.~\ref{Plots}b shows that the maximal adhesion force is reached on
a rigid substrate, $\gamma = 0$, where Eq.~\ref{elasticsubstrate}
reduces to Eq.~\ref{arc}. As the substrate becomes softer, $F$
decreases because the moving adhesion points effectively decrease the
spanning distance $d$. Indeed it is well known that cellular traction
is weaker on softer substrates \cite{c:pelh97}. Like for rigid
substrates the force decreases with increasing opening angle. Both
these trends are also evident in Fig.~\ref{Shapes}a-c, where we show
computed cell shapes for triangles, squares and pentagons on stiff
(red) and elastic (blue) substrates.

\textit{Tension-elasticity model.} Up to now, our results were derived
with the assumption of constant line tension $\lambda$.  Recently it
was suggested that this quantity has an elastic origin, i.e.~$\lambda
= EAu$ where $u=(L-L_0)/L_0$ is the strain induced in the elastic
contour with rigidity $EA$. $L_0= \alpha d$ is the resting length
assumed to be proportional to the spanning distance $d$ with a
dimensionless resting length parameter $\alpha$. The arc contour
length follows from geometrical considerations as $L=2R \arcsin
(d/2R)$, where $R$ is the arc radius.  Hence, the line tension
$\lambda$ itself becomes a function of the adhesion geometry:
\begin{equation}
\label{ElasticLineTension}
\lambda(d) = EA \left[ \frac{2R}{\alpha d} \arcsin \left( \frac{d}{2R} \right) - 1 \right]
\end{equation}
which increases with $d$. Thus also the radius of curvature $R =
\lambda(d)/\sigma$ increases with $d$, as indeed observed
experimentally \cite{uss:bisc08a}. The distance dependence of the
adhesion force now becomes a non-trivial function of the adhesion
distance $d$. On the one hand, the line tension increases with $d$,
thereby increasing the individual forces pulling on the adhesions.  On
the other hand, the arc curvature $\kappa = \sigma / \lambda(d)$
decreases with increasing $d$, leading to less steep pulling
directions and therefore reduced overall force. In Fig.~\ref{Shapes}d
we compare the results of the tension elasticity model (green) to the
constant tension model (red), where all arcs have the same curvature
$\kappa = \sigma / \lambda$.  For the tension-elasticity model, the
two top arcs across the diagonal have a reduced curvature $\kappa$,
because here the spanning distance $d$ is increased by a factor of
$\sqrt{2}$. The total force magnitude slightly decreases due to less
steep pulling directions, although the tension in the individual arcs
actually increases.

\begin{figure}
\includegraphics[width=\columnwidth]{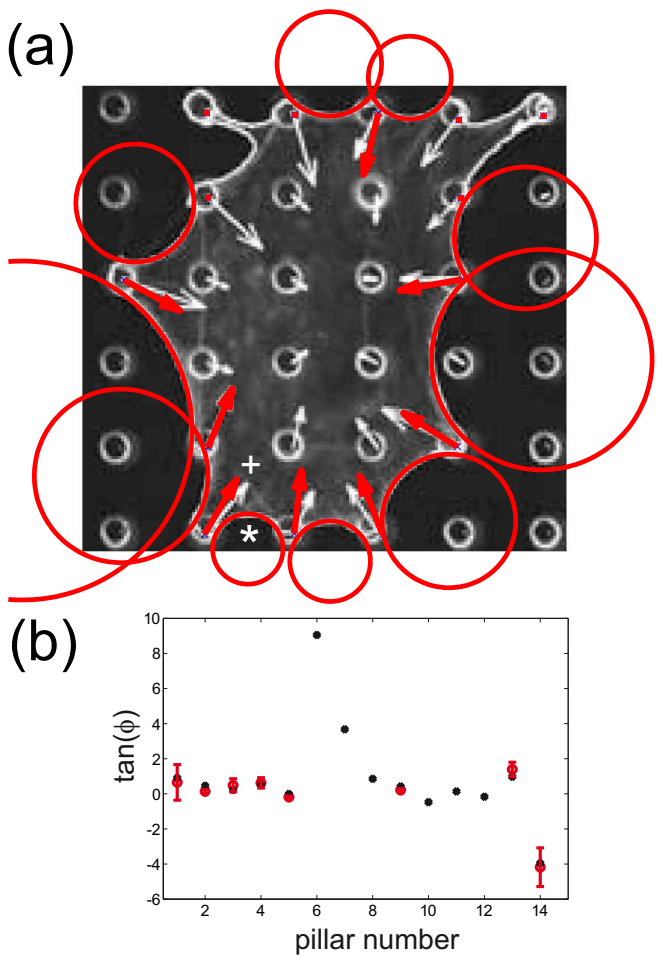}
\caption{\label{Chen} (a) Predicted contour forces (red) and measured
  pillar forces (white) for an endothelial cell cultured on a pillar
  array \cite{c:lemm05}. Force direction is determined by cell shape
  geometry and force magnitude scales with $\sigma$ as determined from
  a least-square fit. Our procedure allows to predict the forces
  within each arc $\lambda=(8,9,15,25,16,9,12,13,38,19)$ nN along the
  contour (counter-clockwise starting at the asterisk). Scales: Pillar
  spacing 9 $\mu$m; force with plus 14 nN. (b) Force directions of
  pillar forces (black) and predicted contour forces (red).}
\end{figure}

\textit{Extracting model parameters from pillar assays.} In general
one expects the surface tension $\sigma$ to be a global quantity and
the line tensions $\lambda$ to be different in different arcs.  Our
model suggests a simple procedure to estimate numerical values from
the shape geometry. Fig.~\ref{Chen}a shows experimental data for an
endothelial cell on a pillar array (Fig.~6 from Ref.~\cite{c:lemm05}).
The contour was fitted by circular arcs using the procedure from
Ref.~\cite{uss:bisc08a}. Because here we focus on contour effects, we
exclude all arcs which might interact with other arcs due to close
proximity or which might be distorted by internal stress fibers.  The
force on a pillar is the vector sum of the two adjacent arc forces.
Assuming constant surface tension, the ratio of arc radii equals the
ratio of line tensions, $R_1/R_2 = \lambda_1/\lambda_2$, and the
respective pulling angles $\beta_i$ are given by $\cos \beta_i = R_i/2
d_i$.  Thus, the resultant \textit{directions} of the contour forces
$\vec F$ are determined by $R$ and $d$ only. Fig.~\ref{Chen}b shows
that we obtain excellent agreement between the predicted
directions of contour forces and the measured directions of the pillar
forces (error bars result from an estimated 10 percent uncertainty).
Because the surface tension determines the \textit{magnitudes} of the
contour forces, we extract a value of $\sigma\approx 2$ nN/$\mu$m by a
least-square fit to the experimental data, which is at the upper limit
of the range of values reported earlier for cortical tension. The
predicted contour forces (red) are shown in Fig.~\ref{Chen}a and
compare favourably with the measured pillar forces (white).

\textit{Discussion.} Our model shows that shape and forces are closely
related in cell adhesion and demonstrates how the spatial distribution
of adhesion sites determines the forces acting at sites of adhesion.
The effect of these forces might be further amplified by feeding into
force or displacement-dependent regulation of cytoskeleton and
adhesion sites, which should be included in future modelling. For
example, Eq.~\ref{elasticsubstrate} predicts that displacement $F/K$
first increases linearly and then saturates as a function of inverse stiffness,
with important consequences for strain homeostasis. In this way,
our model can be used to derive quantitative predictions how cell
behaviour and fate can be steered by adhesion geometry and stiffness.

\begin{acknowledgments}
  We thank Franziska Klein, Dirk Lehnert and Martin Bastmeyer for many
  helpful discussions and Chris Chen and Chris Lemmon for providing
  supplementary data for Fig.~4. This work was supported by the
  Center for Modelling and Simulations in the Biosciences (BIOMS) at
  Heidelberg and by the Karlsruhe Institute of Technology (KIT)
  through its Concept for the Future.
\end{acknowledgments}


\end{document}